\documentclass[usegraphicx,usenatbib,useapjfonts,apj]{emulateapj}
\begin{document}

\sloppy
\newcommand{\vk}{{\bf k}}

\def\plotone#1{\centering \leavevmode
\epsfxsize=\columnwidth \epsfbox{#1}}
\def\lessim{\stackrel{<}{{}_\sim}}

\title[Non-Gaussian halo bias and future galaxy surveys]{Non-Gaussian halo bias and future galaxy surveys}

\author{Carmelita Carbone\altaffilmark{1}, Licia Verde 
\altaffilmark{1,2,3} and Sabino Matarrese\altaffilmark{4}}
\altaffiltext{1}{Institute of Space Sciences (CSIC-IEEC), 
UAB, Barcelona 08193, Spain;carbone@ieec.uab.es}
\altaffiltext{2}{ICREA; verde@ieec.uab.es}
\altaffiltext{3}{Dep. Astrophysical Sciences, Princeton University, 
Ivy Lane, Princeton, USA}
\altaffiltext{4}{Dipartimento di Fisica "G. Galilei", Universit\`a 
degli Studi di Padova and INFN, Sezione di Padova, via Marzolo 8, 
35131, Padova, Italy; sabino.matarrese@pd.infn.it}

\begin{abstract}

We forecast constraints on primordial non-Gaussianity achievable from  
forthcoming surveys by exploiting the scale-dependent halo bias  
introduced on large scales by non-Gaussian initial conditions.
We explore the performance of exploiting both the shape of the  
galaxy power-spectrum on large scales and the cross-correlation of  
galaxies with Cosmic Microwave Background maps through the Integrated  
Sachs Wolfe effect.
We find that future surveys can detect primordial non-Gaussianity of  
the local form with a non-Gaussianity parameter $|f_{\rm NL}|$   
of order unity. This is particularly exciting because, while the simplest  
single-field slow-roll models of inflation predict a primordial $|f_{\rm 
NL}|\ll 1$, this signal sources extra contributions to the effective 
$f_{\rm NL}$ of large-scale structures that are 
expected to be above our predicted detection threshold.
\end{abstract}

\section{Introduction}
Tests of deviation from Gaussian initial conditions offer an important 
window into the very early universe and a powerful test for the mechanism 
which generated primordial perturbations.

While standard single-field models of slow-roll inflation lead to small 
departures from Gaussianity, non-standard scenarios allow for a larger 
level of non-Gaussianity (see, e.g., \citet{BKMR04} 
and references therein).

The standard observables to constrain non-Gaussianity are the Cosmic Microwave 
Background (CMB) and the Large-Scale Structure (LSS) of the universe. 
Traditionally, the most popular method to detect primordial non-Gaussianity 
(NG) has been to measure the bispectrum or the three-point 
function of the CMB \citep{VWHK00, komatsuetal05, yadav}, as the LSS 
bispectrum is sensitive to primordial NG only at high redshift 
\citep{VWHK00, Scocc, sefusattikomatsu, Coor06, PPM07}.

A powerful technique is based on the abundance \citep{MVJ00, VJKM01, 
Loverdeetal07, RB00, RGS00} and clustering \citep{GW86, MLB86, LMV88} 
of rare events such as dark matter density peaks, as they trace the tail 
of the underlying distribution. These theoretical predictions have been 
tested against numerical N-body simulations \citep{KNS07,Grossietal07,DDHS07}.

\citet{DDHS07, MV08} showed that primordial NG affects the clustering of 
dark matter halos inducing a scale-dependent large-scale bias. 
Here we  argue that this effect could be used to constrain NG through 
the Integrated Sachs Wolfe (ISW) effect \citep{isw} and through 
the shape of the galaxy power-spectrum on large scales. 
We forecast how future galaxy surveys could constrain Gaussianity via 
this ``halo-bias" effect. We find that constraints from surveys which 
provide a large sample of galaxies or galaxy clusters over a volume comparable 
to the horizon size (e.g., DES, PanSTARRS, LSST, EUCLID, ADEPT) 
are competitive with CMB bispectrum constraints achievable with an ideal 
CMB experiment. In particular, \citet{bartolofnl05} showed that even when the 
primordial $f_{\rm NL}$ is tiny, the evolution of 
perturbations on super-Hubble scales, yields extra contributions 
to the effective $f_{\rm NL}$ relevant for 
the LSS, which are configuration and redshift dependent. These contributions 
are of amplitude comparable to the forecasted errors and can therefore 
no longer be neglected.

Along the way we offer physical insights in the findings of \cite{MV08} 
(hereafter MV08) and explain the connections to the approach of 
\cite{DDHS07}; we show 
that only the formalism of MV08 can be used to correctly handle 
the non-linear contributions to the primordial NG.

\section{Non-Gaussian halo bias in context}
\citet{MV08} generalized to NG initial conditions 
results and techniques 
developed in the 80s \citep{GW86, MLB86, LMV88} to relate the clustering 
properties of the collapsed structures 
(halos) to those of the underlying dark matter distribution 
for Gaussian 
initial conditions. This approach, which we briefly review here, 
yields an analytic expression for the bias 
of dark matter halos for non-local and scale-dependent NG models in which the 
primordial bispectrum of the potential 
is the dominant higher-order correlation and has a general form.
The starting point of MV08 is to consider the expression for the correlation 
function of regions above a high threshold in the general NG case, which 
has the form \citep{GW86, MLB86, LMV88}:
\begin{equation}
\xi_{h, M}(r)=-1+\exp[X(r)]
\label{eq:X}
\end{equation} 
where $X$ is a complicated expression that depends on all 
the $n$-point correlations of the underlying density field  
filtered on the mass-scale $M$, as well as on the threshold height  
(see MV08 for details).  
Then, one interprets the region above high thresholds as halos, and, for 
large separations $r$ (small values of $X$), expands the exponential to first 
order and considers only terms up to the three--point correlation function. 
Finally, one Fourier transforms to obtain an approximated expression 
for the power-spectrum of halos in NG models.
For NG of the type \citep{SalopekBond90, Ganguietal94,VWHK00, KS01}
\begin{equation}
\Phi=\phi+f_{\rm NL} \left(\phi^2-\langle \phi^2 \rangle\right) \;, 
\label{eq:fnl}
\end{equation}
where $\Phi$ denotes Bardeen's gauge-invariant potential\footnote{On 
sub-Hubble scales, Bardeen's gauge-invariant potential $\Phi$ 
reduces to the usual Newtonian peculiar 
gravitational potential, up to a minus sign. In the literature, 
there are two conventions for Eq.~(\ref{eq:fnl}): the LSS and the CMB one. 
Following the LSS convention, here $\Phi$ is linearly extrapolated at $z=0$. 
In the CMB convention $\Phi$ is instead primordial: thus 
$f_{\rm NL}= (g(z=\infty)/g(0) ) f_{\rm NL}^{CMB}$.} and $\phi$ denotes a 
Gaussian random field, the halo power-spectrum has the form
\begin{equation}
P_{\rm h}(k,z)=\frac{\delta_c^2(z)P_{\delta \delta}(k, z)}{\sigma_M^4 D^2(z)}
\left[ 1+4 f_{\rm NL}\delta_c(z) \alpha(k) \right] \;.   
\label{eq:pkhalobias} 
\end{equation}
Here $\alpha(k)$ is the quantity in Fig. 3 of MV08, which is 
$\propto 1/k^2$ at large scales; $\sigma_M^2$ is the mass variance 
linearly extrapolated to $z=0$; $\delta_c(z) = \Delta_c(z)/D(z)$, with 
$\Delta_c(z)$ the linear overdensity for spherical collapse  
(which weakly depends on $z$ in non-Einstein de Sitter cosmologies),  
and $D(z)$ linear growth-factor of density fluctuations 
normalized to unity at $z=0$.

The Lagrangian bias $b_L$ of the halos (all correlations and peaks considered 
here are those of the {\it initial} density field, linearly extrapolated 
till the present time) is then   
$b_{L,h}(z,M) \simeq b_{L,h}^G(z,M)[1+ 2 f_{\rm NL}\delta_c(z) \alpha(k) ]$,
with $b_{L,h}^G(z,M) \simeq \delta_c(z)/(\sigma_M^2 D(z))$. 
Making the standard assumption that halos move coherently 
with the underlying dark matter, one obtains the 
final Eulerian bias as $b_E=1+b_L$, so that 
\begin{equation}
b_{\rm h} ^{f_{\rm NL}} = 
1+\frac{\delta_c(z_f)}{\sigma_M^2 D(z_o)} \left[ 1+2 f_{\rm NL} \delta_c(z_f)
\alpha(k) \right] \;,
\label{eq:db}
\end{equation}
where, following the approach of \citet{CLMP98}, 
we have made explicit the dependence on both the halo formation redshift 
$z_f$ and the observation redshift $z_o$. 
This expression for the NG bias of halos of mass $M$ is scale-dependent 
and increases rapidly at large scales. 
The approach of identifying peaks with halos is valid for rare (massive) 
halos. Indeed, the above expression for the Gaussian Lagrangian halo bias 
is approximate; a more accurate expression is 
\citep{Eetal88, CK89, MoWhite96}
\begin{equation}
b^G_{L,h}(z_o,M,z_f) = \frac{1}{D(z_o)}\left[\frac{\delta_c(z_f)}{\sigma_M^2}
-\frac{1}{\delta_c(z_f)}\right] \;. 
\label{eq:gaussian_halo_bias_full}
\end{equation} 
In addition, and as discussed below, this expression can be further refined 
with the help of N-body simulations. 
For objects that did not undergo recent mergers, $z_f \gg z_o$,
the bias is well approximated by Eq.~(\ref{eq:db}). Eq.~(\ref{eq:db}) however 
also applies to the case $z_f \approx z_0$ (rapid mergers)  
for $\delta_c^2 \gg \sigma_M^2$, i.e.  
large masses and/or high formation redshifts. 

Note that Eq.~(\ref{eq:db}) can be rewritten as 
$b_{L,h}^{f_{\rm NL}}=b_{L,h}^{G}+\Delta b$, where
\begin{equation}
\label{deltab}
\Delta b = 2 f_{\rm NL} \delta_c(z_f)\left(b^G_{h}(z_o,M,z_f) - 1\right)
\alpha(k)\,.
\end{equation}

One may note that for $f_{\rm NL}$ large and negative, 
Eq.~(\ref{eq:pkhalobias}) would formally yield $b_h^{f_{\rm NL}}$ and 
$P_h(k)$ negative on large enough scales. 
This is a manifestation of the breakdown of the approximations made: 
a) all correlations of higher order than the bispectrum were neglected: 
for large NG this truncation may not hold; b) The exponential in 
Eq.~(\ref{eq:X}) was expanded to linear order. 
This however could be easily corrected for, remembering that the $P(k)$ 
obtained in Eq.~(\ref{eq:pkhalobias}) is in reality the Fourier transform of 
$X$, the  argument of the exponential. 
One would then compute the halo correlation function using Eq.~(\ref{eq:X}) 
and Fourier transforming back to obtain the halo power-spectrum.

\cite{DDHS07} and \cite{slosaretal08} use the peak-background formalism to 
obtain an equation similar to Eq.~(\ref{eq:db}).
In particular, while \cite{DDHS07} relies on the spherical collapse model 
(and thus on the standard Press-Schecter \citep{PS} approach), 
\cite{slosaretal08} extend and reformulate it so that it relies on the 
extended Press-Schecter approach and on the universality of the mass function, 
and can be obtained for any mass function, even one that is a fit to N-body 
simulations. The advantage of their formulation over that of \cite{DDHS07} is 
that they can include the description of the effect of halo mergers. 
The effect of halo merger can be analytically described only for the 
standard Press-Shechter mass function, but they argue that the scaling of 
the correction for mergers could be calibrated from N-body simulations.

We note here that their derivation can {\it only} be carried out for NG of the 
local type. On the other hand, the formulation of MV08 is more 
general: the extended Press-Shechter approach, halo mergers and mass 
functions that are better fit to N-body simulations than the standard 
Press-Shechter can be readily taken into account by substituting 
$b_{L,h}^G$ in Eq.~(\ref{eq:gaussian_halo_bias_full}) by the peak-background 
split  bias which is obtained as 
$b^G_{L,h}=-n^{-1} \partial n/\partial \delta_c$, 
where $n$ denotes the halo mass function for Gaussian initial conditions, 
which could be given e.g. by the \cite{ST} formula or by a 
fit to simulations.

In particular, Eq.~(\ref{eq:gaussian_halo_bias_full}) is replaced by
\begin{equation}
 b^G_{L,h}(z_o,M,z_f)=\frac{1}{D(z_o)}
\left[\frac{q\delta_c(z_f)}{\sigma_M^2}-\frac{1}{\delta_c(z_f)}\right] 
\end{equation}
$$
+ \frac{2p}{\delta_c(z_f)D(z_o)} \left[1 + 
\left(\frac{q\delta_c^2(z_f)}{\sigma_M^2} 
\right)^p\right]^{-1} \;.
$$
The parameters $q$ and $p$ account for non-spherical 
collapse and fit to numerical simulations yield $q\sim 0.75$, $p=0.3$ 
\citep{ST,SMT}. 
The correction for non-spherical collapse also applies to the NG 
correction to the halo bias:
\begin{equation}
\Delta b= 2 q' f_{\rm NL} 
\delta_c (b^G_{h} - 1 )\alpha(k)\,.
\end{equation} 
where $q'$ can be calibrated to N-body simulations and is found to be 
$q'\approx 0.8$ \citep{margotinprep}.
We stress here that it is particularly important to be able to 
account for general non-local and scale-dependent NG characterized 
by a given bispectrum of the potential. Indeed, as shown 
by \cite{bartolofnl05}, there are extra contributions to the bispectrum that 
come in at the same level as the primordial signal: in other words the 
primordial contribution is enhanced for LSS, 
yielding configuration and redshift dependent contributions to the 
{\it effective} $f_{\rm NL}$ which cannot be neglected. Note that the 
NG bias formula of MV08 is fully general and can easily account 
for non-constant $f_{\rm NL}$.

\section{Method}
Before we present the forecasts we need to make two considerations.\\
{\it Halos vs Galaxies}: The theory developed in MV08 
and above describes the clustering properties of halos, 
but we observe galaxies. Different galaxy populations occupy dark matter 
halos following different prescriptions. 
If we think in the halo-model \citep{CooraySheth, SSHJ, PS00, Smithetal03} 
framework, at very large scales (as those relevant for this analysis), 
only the ``two-halo" contribution matters and the details of the  halo 
occupation distribution of galaxies (the so-called ``one-halo" term) 
is unimportant. In particular, the galaxy population known 
as ``luminous red galaxies" (LRG), is known to be old and 
free from recent merger 
activity. For this population, the  modeling of MV08 should offer a 
good description.
Emission lines galaxies on the other hand may be affected by recent merger 
activity and so their $b_h^G$ may need to be modified, as described in \S 2.  
As we discuss below, with this modification, the effect of uncertainties 
in the Gaussian bias enters in our estimates only through the shot-noise 
contribution to the signal-to-noise ratio.\\ 
{\it Detection vs Measurement}:
Before we present the forecasts we need to make an important distinction 
between ``detection" and ``measurement". To compute the statistical 
significance of a detection we need to compute the significance of deviations 
from the null hypothesis: in particular, the fiducial model used in the 
calculation has $f_{\rm NL}=0$ and the error bars are also computed 
assuming the null hypothesis $f_{\rm NL}=0$.
However to carry out a measurement of $f_{\rm NL}\ne 0$, the theoretical model 
and the error bars must be computed as functions of $f_{\rm NL}$. 
Here we will report forecasts for {\it detection} of $f_{\rm NL}$.

We consider two probes: the large-scale power-spectrum of galaxies and 
the ISW effect. In both cases, since we will consider tracers of rare halos, we
 will assume $z_f=z_o$, yielding a possibly conservative estimate of the errors on  $f_{\rm NL}$. 
 
\subsection{Forecasts from the shape of the large-scale power-spectrum}
In the Fisher matrix approach to error forecasts we  can write 
\begin{equation}
\ln {\cal L}=-\frac{1}{2}\frac{(\Delta P(k))^2}{\sigma_P^2}=
-\frac{1}{2}\frac{(P(k)4\delta_c(z)\alpha(k) f_{\rm NL})^2}{\sigma_P^2}
\end{equation}
where we have assumed a Gaussian likelihood \footnote{While strictly 
speaking the distribution of $P(k)$ is non-Gaussian, this is a 
standard assumption in Fisher based approaches.} . 
An estimate of the error in $f_{\rm NL}$, $\sigma_{f_{\rm NL}}(k)$, 
at a given $k$  is given by 
\begin{equation}
\frac{1}{\sigma_{f_{\rm NL}}(k)^2}=\frac{\partial^2 
|\ln {\cal L}|}{\partial f_{\rm NL}^2} = 
\frac{16 P^2\alpha(k)^2\delta_c^2}{\sigma_P^2}
\end{equation}
The relative error in $P$ for a shell in $k$-space of width 
$\Delta k$ and for a survey with effective volume $V_{eff}$ is
\begin{equation}
\left( \frac{\sigma_P}{P}\right)^2= 
\frac{2}{4 \pi k^2 \Delta k V_{eff}/(2\pi^3)}\,,
\end{equation}
where $V_{eff}=V(1+1/(\bar{n}P))$, with $V$ the survey volume, 
and $\bar{n}$  the average density of galaxies. 
If shot-noise is sub-dominant (i.e. $\bar{n}P\gg1$) $V_{eff}=V$.

The  total  error from a $k$-range from $k_{min}$ to $k_{max}$ is:
\begin{equation}
\frac{1}{\sigma_{f_{\rm NL}}^2} = 
\frac{8}{2\pi^2}\frac{\Delta_c(z)^2}{D(z)^2} V_{eff} 
\int_{k_{min}}^{k_{max}} \alpha(k)^2k^2 dk\,.
\end{equation}
If we divide the survey in redshift slices centered around $z_i$ 
then the errors  obtained combining different redshift slices 
(if uncorrelated) is 
\begin{equation}
\frac{1}{\sigma_{f_{\rm NL}}^2}=\sum_i\frac{1}{\sigma_{f_{\rm NL}}(z_i)^2}\,.
\end{equation}
We set $k_{max}$ to be  $0.03$ $h$/Mpc and $k_{min}$ to be greater than 
the $2\pi/V^{1/3}$ where $V$ is the volume of the shell considered 
($\Delta z=0.1$).
Note that the effect of NG alters the broad-band behavior of the $P(k)$ 
on very large scales,  which is not affected  by the precision with 
which the radial positions of the galaxies in measure. 
Thus, we can treat photometric and spectroscopic surveys on 
the same footing.
The requirement of surveying a large volume of the universe and sampling
highly biased galaxies to beat shot-noise, which is a key point for 
BAO surveys is also a bonus for constraining primordial NG. 
In particular, for the $k_{min}$ we use we find that 
$P_g(k_{min})\simeq P_g(k=0.2 h{\rm/Mpc})$, thus the shot-noise 
requirement for BAO surveys of $\bar{n}P(k=0.2{\rm h/Mpc}) >  1$ 
implies that for all scales of interest here $\bar{n}P\gg 1$. 
We have checked that our results do not change if we impose $\bar{n}P \sim 3$.

While for BAO surveys  measuring redshift accurately is crucial 
\citep{SeoEisenstein03, BlakeBridle05}, for this application is not important.
\begin{deluxetable*}{lcccccc}
\tablecolumns{7}
\tablecaption{Galaxy Surveys considered}
\tablehead{\colhead{survey} & \colhead{z range} & \colhead{sq deg} &\colhead{mean galaxy density $(h/Mpc)^3$} & \colhead{$\Delta f_{\rm NL}/q'$ LSS} }
\startdata
SDSS LRG's &  $0.16<z<0.47$  &  $7.6 \times 10^3$  &  $1.36\times 10^{-4}$  &  $40$  \nl
BOSS     &  $0<z<0.7$  &  $10^4$  &  $2.66\times 10^{-4}$  &  $18$ \nl
WFMOS low z &  $0.5<z<1.3$  &  $2 \times 10^3$  &  $4.88\times 10^{-4}$  &  $15$ \nl
WFMOS high z &  $2.3<z<3.3$  &  $3 \times 10^2$  &  $4.55 \times 10^{-4}$  &  $17$ \nl
ADEPT  &  $1<z<2$  &  $2.8 \times 10^4$  &  $9.37 \times 10^{-4}$  &  $1.5$ \nl
EUCLID &  $0<z<2$  &  $2 \times 10^4$  &  $1.56 \times 10^{-3}$  &  $1.7$ \nl
DES &  $0.2<z<1.3$  &  $5 \times 10^3$&$1.85 \times 10^{-3}$  &  $8$ \nl
PanSTARRS  &  $0<z<1.2$  &  $3 \times 10^4$  &  $1.72 \times 10^{-3}$  &  $3.5$ \nl
LSST    &  $0.3<z<3.6$  &  $3 \times 10^4$  &  $2.77 \times 10^{-3}$ & $0.7$
\enddata
\label{tab:surveys}
\end{deluxetable*}

\subsection{Forecasts from the Integrated Sachs Wolfe effect}
The ISW effect probes the largest cosmological 
scales. As the NG effect goes $\propto 1/k^2$ on large scales, this is a 
promising probe of NG. 
Here we follow \citet{Afshordi04} to quantify the significance of a 
detection of $f_{\rm NL}$ through the estimate of the cross-correlation between the ISW 
effect with LRG galaxy distribution.
For a galaxy survey with the average comoving density 
distribution $n_c(r)$ as a function of the comoving distance $r$,
in the Limber approximation, the expected cross-correlation in the spherical 
harmonic space, can be written as
\begin{equation}
\label{Cgt}
C_{gT} (\ell) 
=\frac{2~ T}{\int dr ~r^2 n_c(r)}\int dr ~n_c(r) P_{\Phi^{\prime},g}(k)\, ,
\end{equation}
where $k=(\ell+1/2)/r$,  $\Phi^{\prime}$  is the derivative
of the gravitational potential with respect to the conformal time, and 
$\delta^{\rm 2D}_{g, \ell m}$ and
$T_{\ell m}$ are the projected survey galaxy overdensity and the CMB
temperature in the spherical harmonic space, respectively.

The expected dispersion in the cross-correlation signal is $
\Delta C^2_{gT}(\ell) \simeq C_{gg}(\ell) C_{TT}(\ell) [f_{\rm sky}(2\ell+1)]^{-1}$, 
where $f_{\rm sky}$ is the fraction of sky covered in the survey, and
we assumed a small cross-correlation signal, i.e.
$C^2_{gT}(\ell) \ll C_{gg}(\ell) C_{TT}(\ell)$.

For a galaxy distribution  biased according 
to Eq.~(\ref{eq:db}), dividing the survey in redshift shells, and 
following the same procedure of Section~3, 
the error  in each shell at redshift $z$  for a given $\ell$ is 
\begin{equation}
\sigma_{f_{\rm NL}}^{-2}= \frac{\gamma \big[H(z)D(z)\frac{d}{dz}((1+z)D(z))P_{\delta \delta}(k,0) \Delta b(k,z)\big]^2\! r^2 \!\delta r}{(2l+1)^3C_{TT}(\ell)[P_G(k,z)+n_c(r)^{-1}]}\,,
\label{sigmafnl}
\end{equation}
where $\gamma=8 f_{\rm sky}\left(3TH_0^2\Omega_{m0}/c^3 \right)^2$, 
$k\equiv(l+1/2)/r$, $\delta r= (c/H(z))\Delta z$, $\Delta b$ is 
Eq.~(\ref{deltab}) in the limit $f_{\rm NL}=1$, and 
$P_G$ denotes the galaxy power spectrum in the Gaussian case. 
We impose $k_{min}$ to be greater than the largest mode that can be sampled 
in each survey shell and $k_{max}=0.03$ $h$/Mpc. The total error is obtained summing up 
Eq.~(\ref{sigmafnl}) on all the multipoles $\ell \le 200$
and integrating over the minimum and maximum redshift
of each survey.

For future large-scale galaxy surveys, we obtain 
$\Delta f_{\rm NL}=7.6, 12.5, 11.5$ for LSST, EUCLID and ADEPT, respectively. 

\section{Results \& discussion}
Here we present forecasts of $f_{\rm NL}$  constraints for forthcoming and 
future surveys. The surveys we consider and their specifications are 
reported in Table \ref{tab:surveys}, along with the $1-\sigma$ error on 
$f_{\rm NL}$ from the shape of the galaxy power-spectrum. 
The reported errors on $f_{\rm NL}$ have been normalized by the 
correction factor for non-spherical collapse $q'\approx 0.8$. 
Note that the number of galaxies and the Gaussian bias enter in this 
signal-to-noise calculation only through the contribution to the error due to 
shot-noise. The reported numbers are not dominated by shot-noise.  

This signal-to-noise calculation indicates that the halo clustering
approach to primordial NG is in principle more promising
than the ISW one: the ISW signal is weighted at low redhift ($z \lessim 1$),
when dark energy dominates, while the effect of 
NG grows with redshift. However, the two approaches  are
affected by different systematics and thus should be considered
complementary. 

It is interesting to compare the constraints on primordial NG achievable 
from the large-scale halo clustering with those achievable with the 
small-scale galaxy bispectrum. For example, comparing with 
\citet{sefusattikomatsu} we deduce that the halo-clustering constraints are 
a factor of $3$ stronger than the bispectrum ones. 
The bispectrum however, through its dependence on the $k$-space
configuration, can be used to discriminate among different forms of NG. 
The CMB bispectrum for an ideal experiment can yield constraints of 
$\Delta f_{\rm NL}=$ few \citep{YKW07}.
The results  of Table \ref{tab:surveys} indicate that constraints on 
$f_{\rm NL}$ of order unity are achievable with future surveys, making 
it a highly competitive technique.  
We conclude that it is particularly important to be able to take into account 
general non-local and scale-dependent NG features characterized by a given 
bispectrum of the potential. In fact, as shown by \cite{bartolofnl05}, there 
are contributions to the bispectrum, which have a specific shape and 
redshift dependence and which come into play at the level of 
$f_{\rm NL}\sim$ few. This is well above the detection threshold 
for forthcoming and proposed surveys, thus opening up the possibility 
to measure these secondary contributions to $f_{\rm NL}$. \\ 

While this work was being completed we became aware of Afshordi \& Tolley 
(arxiv:0806.1061) and of McDonald (arxiv:0806.1046). 
Our results are in good agreement with theirs.\\

{\it Acknowledgments: }
CC is supported through a Beatrix de Pinos grant.
LV is supported by FP7-PEOPLE-2007-4-3-IRG n. 202182 and CSIC I3 grant 
n. 200750I034. LV  thanks W. Hu, M. Grossi and E. Branchini for fruitful discussions. 
SM acknowledges partial support by ASI contract I/016/07/0 "COFIS" 
and ASI contract Planck LFI Activity of Phase E2.


\end{document}